\documentclass[11pt,a4paper]{article}
\pdfoutput=1
\usepackage{jheppub}

\usepackage{url}
\usepackage{bm}

\allowdisplaybreaks[2]

\def\cA{\mathcal{A}}
\def\cB{\mathcal{B}}

\def\cD{\mathcal{D}}
\def\cE{\mathcal{E}}

\def\cN{\mathcal{N}}
\def\cO{\mathcal{O}}
\def\cP{\mathcal{P}}

\def\mint{\int_{-\infty}^\infty\!\cdots\!\int_{-\infty}^\infty}

\newcommand{\be}{\begin{equation}}
\newcommand{\ee}{\end{equation}}
\newcommand{\ba}{\begin{aligned}}
\newcommand{\ea}{\end{aligned}}

\def\({\left(}
\def\){\right)}

\newcommand{\del}{\partial}

\newcommand{\re}{{\rm e}}

\newcommand{\rd}{{\rm d}}

\newcommand{\x}{\mathsf{x}}
\newcommand{\y}{\mathsf{y}}

\preprint{RUP-20-7, USTC-ICTS/PCFT-20-10}

\title{Bloch electrons on honeycomb lattice and toric Calabi-Yau geometry}

\author[a]{Yasuyuki Hatsuda}
\author[b,c]{and Yuji Sugimoto}

\affiliation[a]{Department of Physics, Rikkyo University,\\Toshima, Tokyo 171-8501, Japan}
\affiliation[b]{Interdisciplinary Center for Theoretical  Study, University of Science and Technology of China, Hefei, Anhui 230026, China}
\affiliation[c]{Peng Huanwu Center for Fundamental Theory, Hefei, Anhui 230026, China}

\abstract{
We find a new relation between the spectral problem for Bloch electrons on a two-dimensional honeycomb lattice in a uniform magnetic field and that for quantum geometry of a toric Calabi-Yau threefold. We show that a difference equation for the Bloch electron is identical to a quantum mirror curve of the Calabi-Yau threefold. As an application, we show that bandwidths of the electron spectra in the weak magnetic flux regime are systematically calculated by the topological string free energies at conifold singular points in the Nekrasov-Shatashvili limit.
}

\begin{document}

\maketitle

\renewcommand{\thefootnote}{\arabic{footnote}}
\setcounter{footnote}{0}
\setcounter{section}{0}

\section{Introduction}

Recently, a new relation between the Hofstadter model \cite{hofstadter1976} and 
the quantum geometry of a toric Calabi-Yau threefold was found \cite{hatsuda2016}. 
The Hofstadter model is a simple two-dimensional square lattice model for Bloch electrons in a uniform magnetic field. 
The electron spectrum in the Hofstadter model shows a remarkably rich behavior.
The basic idea in \cite{hatsuda2016} is to identify the eigenvalue equation for the electron with the quantization of the mirror curve of the toric Calabi-Yau threefold. 
The magnetic flux plays the role of a quantum deformation parameter.

An interesting implication of this correspondence is that the moduli space of the quantum Calabi-Yau geometry seems very complicated. In fact, it was observed in \cite{hatsuda2016} that conifold and orbifold singular points in the moduli space correspond to band edges and van Hove singularities of sub-bands of the electron spectrum, respectively. The structure of sub-bands in the Hofstadter model is fractal for rational magnetic fluxes and the Cantor set for irrational fluxes \cite{hofstadter1976}. 

\begin{figure}[t]
\begin{center}
    \includegraphics[width=0.4\linewidth]{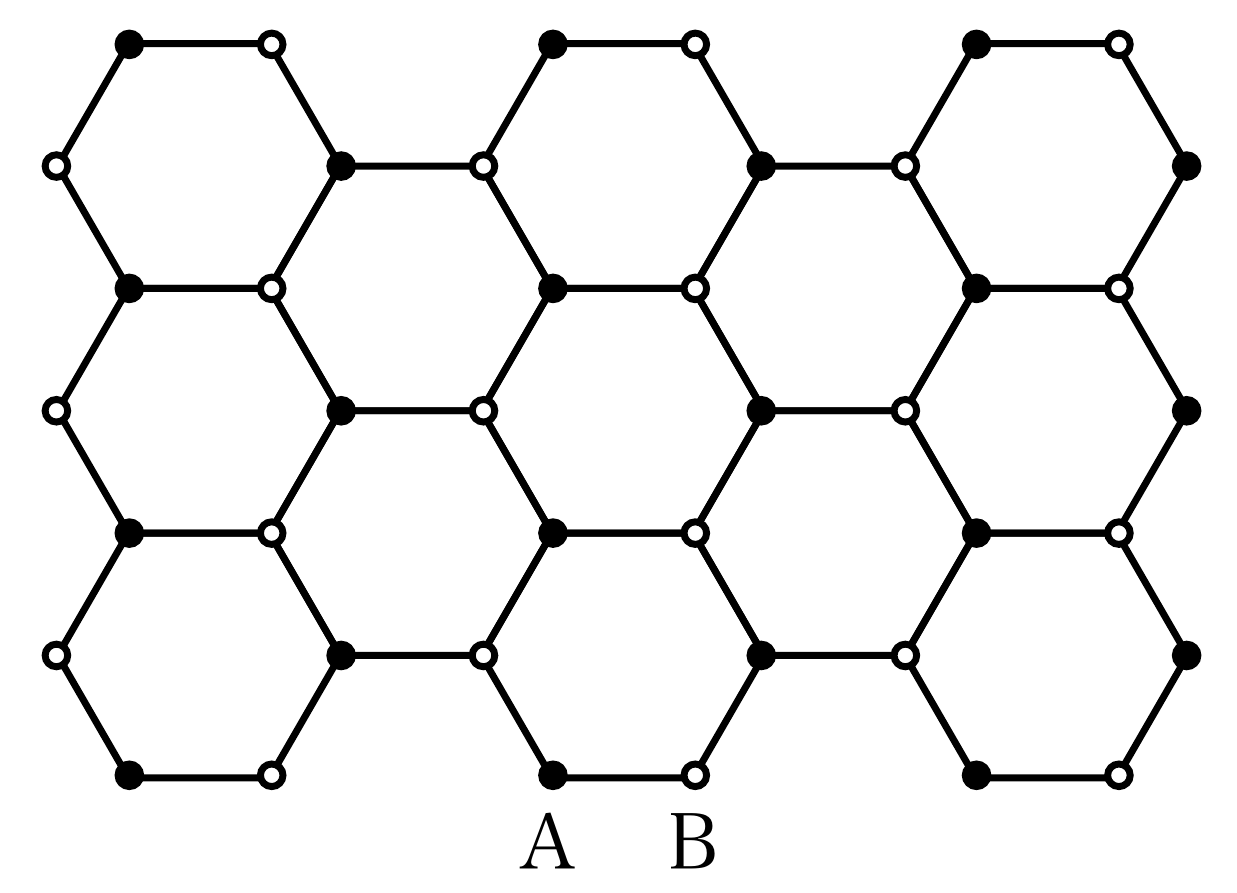}
\end{center}
\vspace{-0.5truecm}
  \caption{The bipartite honeycomb lattice has two kinds of sub-lattices A and B.}
  \label{fig:honeycomb}
\end{figure}

The correspondence was easily generalized to the triangular lattice system \cite{hatsuda2017}.
However, it is far from obvious to extend it to the honeycomb lattice system
because the honeycomb lattice has two sub-lattices as shown in figure~\ref{fig:honeycomb}.
In this paper, we specify a counterpart of the Bloch electron system on the honeycomb lattice.
We show an equivalence between a difference equation obtained from the eigenvalue equations for the electron and a quantized algebraic curve of a toric Calabi-Yau threefold. 
This manifold is identified as local $\cB_3$ in the literature.
Interestingly, the identified geometry is the same as the triangular case in \cite{hatsuda2017}.
The difference comes from moduli parameters in both cases. 
We find that the honeycomb lattice system corresponds to an unconventional moduli identification
while the triangular lattice to a more natural one.

The honeycomb lattice is realized in graphene.
The spectrum of Bloch electrons on the honeycomb lattice in a uniform magnetic field was first studied in \cite{rammal1985}.
Recently, the non-perturbative bandwidth in the weak magnetic flux regime was analyzed in great detail \cite{hatsuda2018}.
However, the analysis in \cite{hatsuda2018} heavily relies on the numerical analysis.
The analytic treatment is lacking.
In this paper, we fill this gap by using the connection with the quantum Calabi-Yau geometry.
We relate the non-perturbative bandwidth to the topological string free energy.
We have powerful techniques to compute this free energy systematically.
As a result, we can predict the bandwidth of the Bloch electron on the honeycomb lattice.
This is a nice application of topological string theory to real physics.

The organization of this paper is as follows.
In section~\ref{sec:Bloch/CY}, we discuss the relation between the Bloch electrons on the honeycomb lattice and the toric Calabi-Yau threefold.
We specify the corresponding geometry.
In section~\ref{sec:application}, we apply this relation to the energy spectrum. 
We show that the bandwidth in the weak magnetic flux regime is computed by the topological string free energy.
Section~\ref{sec:con} is devoted to concluding remarks.
In appendix~\ref{RefTopStr}, we briefly explain how to compute the topological string free energy.

\section{From Bloch electrons to quantum Calabi-Yau geometry}\label{sec:Bloch/CY}
\subsection{Bloch electrons in a honeycomb lattice}\label{sec:honeycomb}
We start with a short review of an electron system in a two-dimensional honeycomb lattice.
We follow the notation in \cite{hatsuda2018}.
The main difference from the Hofstadter model is that the honeycomb lattice is a bipartite system with two sub-lattices. 
We have to treat these sub-lattices separately.
We denote the two sub-lattices as A and B, as shown in figure~\ref{fig:honeycomb}.

We turn on a magnetic field perpendicular to the lattice plane.
As seen in \cite{hatsuda2018}, the eigenvalue equations of the electron are then given by the following two-dimensional difference equations:
\be
\ba
E \Psi_\text{A}(x,y)&=e^{-\frac{i\phi}{3a}x+\frac{i\phi}{12}}\Psi_\text{B}\biggl( x-\frac{a}{2}, y+\frac{\sqrt{3}a}{2} \biggr)
+e^{\frac{i\phi}{3a}x-\frac{i\phi}{12}}\Psi_\text{B}\biggl( x-\frac{a}{2}, y-\frac{\sqrt{3}a}{2} \biggr)
+\Psi_\text{B}(x+a,y), \\
E \Psi_\text{B}(x,y)&=e^{\frac{i\phi}{3a}x+\frac{i\phi}{12}}\Psi_\text{A}\biggl( x+\frac{a}{2}, y-\frac{\sqrt{3}a}{2} \biggr)
+e^{-\frac{i\phi}{3a}x-\frac{i\phi}{12}}\Psi_\text{A}\biggl( x+\frac{a}{2}, y+\frac{\sqrt{3}a}{2} \biggr)
+\Psi_\text{A}(x-a,y).
\ea
\ee
where $a$ is the lattice spacing. 
The magnetic flux $\phi$ is normalized as $\phi=2\pi \Phi/\Phi_0$ where $\Phi$ is the flux per unit cell and $\Phi_0=hc/e$.

These eigenvalue equations are our starting point. 
Since there is no $y$-dependence of the coefficients, we can take the plane wave solution by $\Psi_X(x,y)=e^{ik_y y}\psi_X(x)$, and the eigenvalue problem reduces to the one-dimensional problem:
\be
\ba
E \psi_\text{A}(x)&=2\cos\biggl( \frac{\phi}{3a}x-\frac{\phi}{12}-\frac{\sqrt{3}a}{2}k_y \biggr)
\psi_\text{B}\biggl( x-\frac{a}{2} \biggr)
+\psi_\text{B}(x+a), \\
E \psi_\text{B}(x)&=2\cos\biggl( \frac{\phi}{3a}x+\frac{\phi}{12}-\frac{\sqrt{3}a}{2}k_y \biggr)
\psi_\text{A}\biggl( x+\frac{a}{2} \biggr)
+\psi_\text{A}(x-a).
\ea
\label{eq:diff-0} 
\ee
We can easily eliminate one of these unknown functions.
By eliminating $\psi_\text{A}(x)$, one gets the difference equation for only $\psi_\text{B}(x)$:
\be
\ba
&\lambda \psi_\text{B}(x)=2\cos \biggl( \frac{\phi}{3a}x+\frac{\phi}{12}-\frac{\sqrt{3}a}{2}k_y\biggr) \psi_\text{B}\(x+\frac{3a}{2}\), \\
&+2\cos \biggl( \frac{\phi}{3a}x-\frac{5\phi}{12}-\frac{\sqrt{3}a}{2}k_y\biggr) \psi_\text{B}\(x-\frac{3a}{2}\)
+2\cos \biggl( \frac{2\phi}{3a}x+\frac{\phi}{6}-\sqrt{3}ak_y \biggr) \psi_\text{B}(x),
\ea
\label{eq:diff-1}
\ee
where $\lambda:=E^2-3$.

\begin{figure}[t]
\begin{center}
  \begin{minipage}[b]{0.4\linewidth}
    \centering
    \includegraphics[height=6cm]{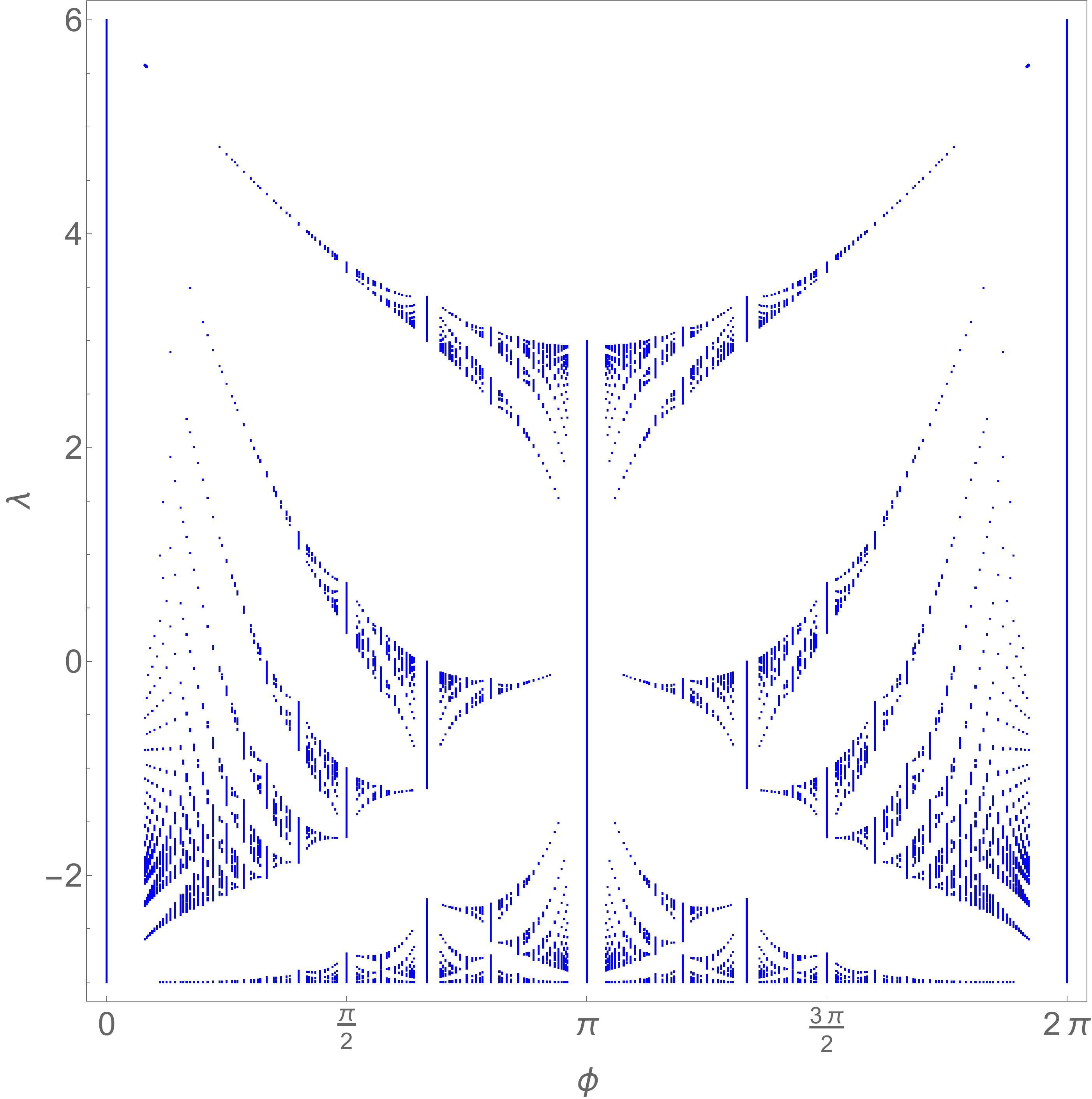}
  \end{minipage} \hspace{1cm}
  \begin{minipage}[b]{0.4\linewidth}
    \centering
    \includegraphics[height=6cm]{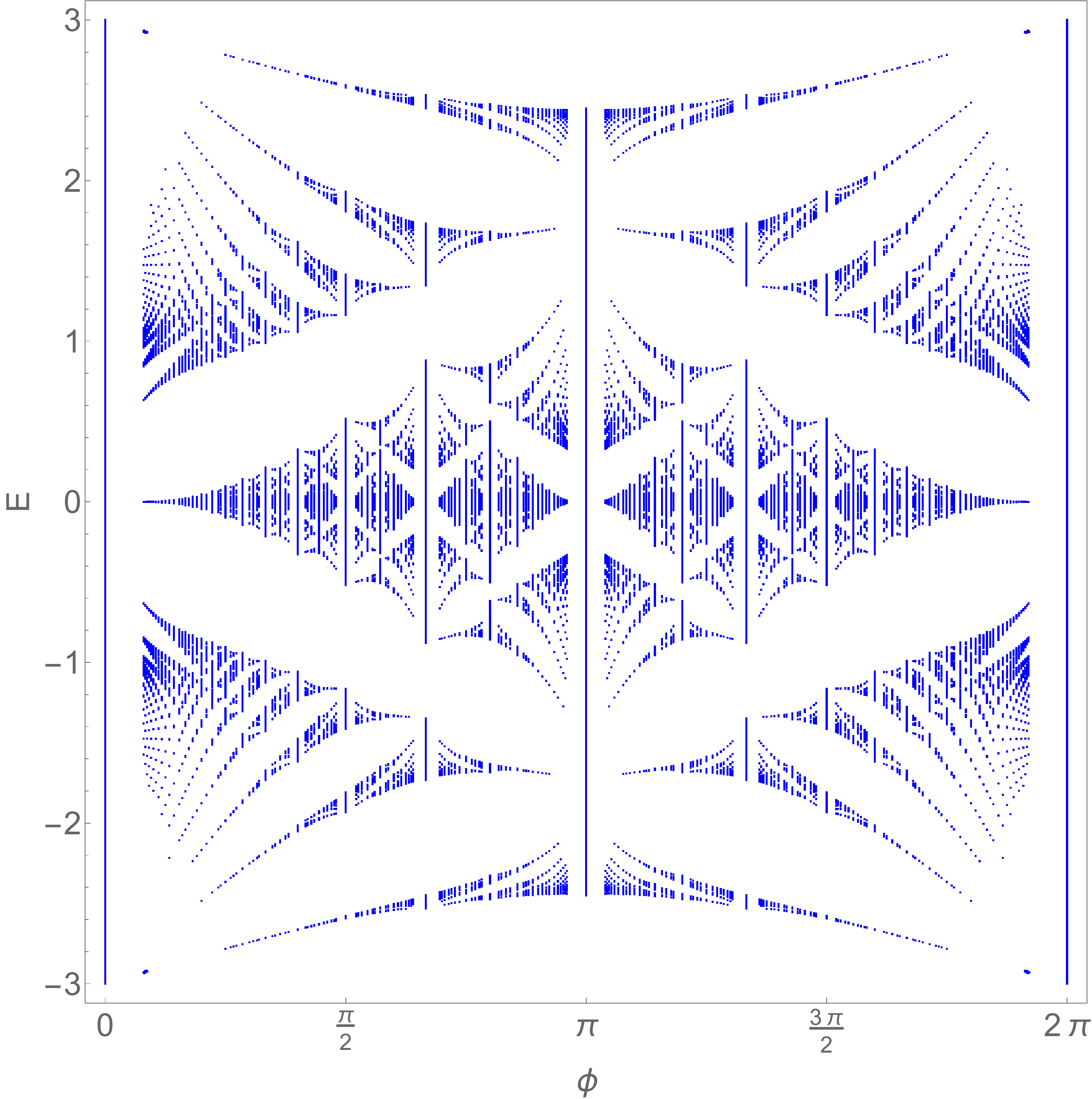}
  \end{minipage} 
\end{center}
  \caption{The spectra of $\lambda$ (left) and of $E$ (right) as functions of the ``rational'' magnetic flux $\phi=2\pi a/b$ with all coprime integers $1\leq a < b \leq 25$.}
  \label{fig:butterfly}
\end{figure}

If the magnetic field is turned off ($\phi=0$), the difference equation leads to
\begin{equation}
\begin{aligned}
\lambda \psi_\text{B}(x)=2\cos \biggl( \frac{\sqrt{3}ak_y}{2}\biggr) \( \psi_\text{B}\(x+\frac{3a}{2}\)+\psi_\text{B}\(x-\frac{3a}{2}\)\)+2\cos( \sqrt{3}ak_y ) \psi_\text{B}(x),
\end{aligned}
\end{equation}
Setting $\psi_\text{B}(x)=e^{ik_x x}$, we obtain the well-known dispersion relation for the honeycomb lattice:
\begin{equation}
\begin{aligned}
E^2=3+4\cos\(\frac{3ak_x}{2} \) \cos \( \frac{\sqrt{3}ak_y}{2}\)+2\cos ( \sqrt{3}ak_y),
\end{aligned}
\end{equation}
where $E=0$ is the zero-gap energy.
For generic $\phi$, the eigenvalue of \eqref{eq:diff-0} or \eqref{eq:diff-1} is quite rich.
We show the spectra of $\lambda$ and of $E$ as functions of rational $\phi$ in figure~\ref{fig:butterfly}.

\subsection{Identifying the toric Calabi-Yau geometry}
In this subsection, we look for the quantum mirror geometry of a toric Calabi-Yau threefold corresponding to the Bloch electron on the honeycomb lattice.
%The toric Calabi-Yau we focus on is so-called local $\cB_3$ geometry.
Originally, the eigenvalue equations for the honeycomb lattice system are given by a couple of equations \eqref{eq:diff-0}.
In this picture, it is not easy to directly identify a counterpart on the Calabi-Yau side.
However, once we use the reduced equation \eqref{eq:diff-1}, we can find the corresponding geometry.
Our conclusion is that the local $\cB_3$ geometry with unconventional moduli parameters describes the honeycomb lattice system.
Note that the same geometry also describes the triangular lattice \cite{hatsuda2017}. As we will explain below, the difference is just the moduli identification.

For this purpose, we first shift the argument of \eqref{eq:diff-1} 
\be
x \to x+\frac{a}{2}+\frac{3\sqrt{3}a^2}{2\phi}k_y, \qquad
\psi(x):=\psi_\text{B}\biggl( x+\frac{a}{2}+\frac{3\sqrt{3}a^2}{2\phi}k_y \biggr).
\ee
Then the difference equation \eqref{eq:diff-1} leads to the symmetric form
\be\ba
\lambda \psi(x)=2\cos\( \frac{\phi}{3a}x+\frac{\phi}{4} \)\psi\(x+\frac{3a}{2} \)
+2\cos\( \frac{\phi}{3a}x-\frac{\phi}{4} \) \psi\( x-\frac{3a}{2} \) \\
+2\cos \( \frac{2\phi}{3a}x+\frac{\phi}{2} \) \psi(x).
\ea
\label{eq:diff-2}
\ee
We define the Hamiltonian operator on the right hand side by
\be
H=2\cos\( \frac{\phi}{3a}x+\frac{\phi}{4} \) e^{\frac{3a}{2}\del_x}
+2\cos\( \frac{\phi}{3a}x-\frac{\phi}{4} \)e^{-\frac{3a}{2}\del_x}
+2\cos \( \frac{2\phi}{3a}x+\frac{\phi}{2} \)
\ee
Let us further introduce new canonical operators by
\be
q:=\frac{\phi}{3a}x,\qquad p:=\frac{3a}{i}\del_x,\qquad [q,p]=i\phi.
\ee
Then one finds
\be
\ba
H&=e^{\frac{i\phi}{4}}e^{iq}e^{\frac{ip}{2}}+e^{-\frac{i\phi}{4}}e^{-iq}e^{\frac{ip}{2}}
+e^{-\frac{i\phi}{4}}e^{iq}e^{-\frac{ip}{2}}+e^{\frac{i\phi}{4}}e^{-iq}e^{-\frac{ip}{2}}
+e^{\frac{i\phi}{2}}e^{2iq}+e^{-\frac{i\phi}{2}}e^{-2iq} \\
&=e^{iq+\frac{ip}{2}}+e^{-iq+\frac{ip}{2}}+e^{iq-\frac{ip}{2}}+e^{-iq-\frac{ip}{2}}+e^{\frac{i\phi}{2}}e^{2iq}+e^{-\frac{i\phi}{2}}e^{-2iq},
\ea
\label{eq:H-1}
\ee
where we have used the Baker-Campbell-Hausdorff formula.
We finally do the canonical transformation $\x=q-p/2$ and $\y=q+p/2$, and obtain
\be
H=e^{i\x}+e^{-i\x}+e^{i\y}+e^{-i\y}+e^{\frac{i\phi}{2}}e^{i\x+i\y}+e^{-\frac{i\phi}{2}} e^{-i\x-i\y}, \qquad [\x,\y]=i\phi.
\label{eq:H-honey}
\ee
Clearly, the magnetic flux plays the role of the Planck constant in our convention.

Na\"ively, this operator is a quantized operator of the algebraic curve
\be
\cE=e^{ix}+e^{-ix}+e^{iy}+e^{-iy}+e^{ix+iy}+e^{-ix-iy}, \qquad (x,y) \in \mathbb{C}^2.
\label{eq:naive-curve}
\ee
%This curve is known as the mirror curve of local $\cB_3$ geometry.
However, one has to be careful in quantization prescriptions. We follow the quantization scheme in \cite{grassi2016}, where
a classical term $e^{ax+by}$ is quantized by
\begin{equation}
\begin{aligned}
e^{ax+by} \to e^{a\x+b\y}.
\end{aligned}
\label{eq:rule}
\end{equation}
%where $\x$ and $\y$ are quantized operators of $x$ and $y$, respectively.
In this rule, the quantization of the curve \eqref{eq:naive-curve} rather yields
\begin{equation}
\begin{aligned}
H=e^{i\x}+e^{-i\x}+e^{i\y}+e^{-i\y}+e^{i\x+i\y}+e^{-i\x-i\y}.
\end{aligned}
\end{equation}
Note that this operator is just the quantum Hamiltonian for the triangular lattice studied in \cite{hatsuda2017}.
For our purpose, we have to start with the ``classical mirror curve''
\begin{equation}
\begin{aligned}
\cE=e^{ix}+e^{-ix}+e^{iy}+e^{-iy}+e^{\frac{i\phi}{2}}e^{ix+iy}+e^{-\frac{i\phi}{2}}e^{-ix-iy},
\end{aligned}
\label{eq:honeycomb-curve}
\end{equation}
which actually leads to the quantum Hamiltonian \eqref{eq:H-honey} under the quantization rule \eqref{eq:rule} with the identification $\cE=\lambda$.

Let us compare this result with the mirror curve of local $\cB_3$.
The generic form of the mirror curve for this geometry is
\begin{equation}
\begin{aligned}
\cE'=e^{ix}+e^{iy}+e^{-ix-iy}+m_1 e^{-ix}+m_2 e^{-iy}+m_3 e^{ix+iy},
\label{eq:mirror-B3}
\end{aligned}
\end{equation}
where $m_1$, $m_2$, $m_3$ (and $\cE'$) are complex moduli parameters of the mirror geometry.
To rewrite it as a more symmetric form, we shift the variables $x \to x+\frac{1}{2}\log m_1$ and $y \to y+\frac{1}{2}\log m_2$.
Then the mirror curve becomes
\be
\cE'=\sqrt{m_1}(e^{ix}+e^{-ix})+\sqrt{m_2}(e^{iy}+e^{-iy})+\sqrt{m_1 m_2} m_3 e^{ix+iy}+\frac{1}{\sqrt{m_1 m_2}} e^{-ix-iy}.
\label{eq:mirror-B2-another}
\ee
Now setting $m_1=m_2=m_3=e^{i\phi/3}$ and $\cE'=e^{i\phi/6}\cE$, the two curves \eqref{eq:honeycomb-curve} and \eqref{eq:mirror-B2-another} get identical.

We conclude that the spectral problem for the Bloch electron on the honeycomb lattice is equivalent to the quantization of the local $\cB_3$ mirror curve \eqref{eq:mirror-B2-another} under the identification:
\begin{equation}
\begin{aligned}
m_1=m_2=m_3=e^{i\phi/3},\qquad \cE'=e^{i\phi/6}\lambda.
\end{aligned}
\label{eq:mass-honey}
\end{equation}
This identification of the moduli parameters is quite unconventional because the moduli parameters depend on the quantum parameter $\phi$. The classical limit of the quantum geometry is obscure. Nevertheless, we will show in the next section, the topological string theory on this unconventional quantum geometry precisely describes the bandwidth of the electron spectrum on the honeycomb lattice.
Note again that the same geometry also describes the triangular lattice if the moduli are set to $m_1=m_2=m_3=1$.

%%%%%%%%%%%%%%%%%
\paragraph{Asymmetric hopping case}
\par~\\
So far, we have considered the symmetric hopping case. It is easy to generalize it to the asymmetric hopping case.
In this case, the eigenvalue equations we should start with are 
\be
\ba
E \Psi_\text{A}(x,y)&=t_1 e^{-\frac{i\phi}{3a}x+\frac{i\phi}{12}}\Psi_\text{B}\biggl( x-\frac{a}{2}, y+\frac{\sqrt{3}a}{2} \biggr)\\
&\quad+t_2 e^{\frac{i\phi}{3a}x-\frac{i\phi}{12}}\Psi_\text{B}\biggl( x-\frac{a}{2}, y-\frac{\sqrt{3}a}{2} \biggr)
+t_3 \Psi_\text{B}(x+a,y), \\
E \Psi_\text{B}(x,y)&=t_1 e^{\frac{i\phi}{3a}x+\frac{i\phi}{12}}\Psi_\text{A}\biggl( x+\frac{a}{2}, y-\frac{\sqrt{3}a}{2} \biggr)\\
&\quad+t_2 e^{-\frac{i\phi}{3a}x-\frac{i\phi}{12}}\Psi_\text{A}\biggl( x+\frac{a}{2}, y+\frac{\sqrt{3}a}{2} \biggr)
+t_3 \Psi_\text{A}(x-a,y).
\ea
\ee
Repeating the same computation above, the reduced eigenvalue equation \eqref{eq:diff-2} is modified as
\begin{equation}
\begin{aligned}
&\lambda \psi(x)=\( t_1 t_3 e^{i(\frac{\phi x}{3a}+\frac{\phi}{4})}+t_2 t_3 e^{-i(\frac{\phi x}{3a}+\frac{\phi}{4})} \) \psi\( x+\frac{3a}{2}\)\\
&\quad+\( t_1 t_3 e^{-i(\frac{\phi x}{3a}-\frac{\phi}{4})}+t_2 t_3 e^{i(\frac{\phi x}{3a}-\frac{\phi}{4})} \) \psi\( x-\frac{3a}{2}\)
+2t_1 t_2 \cos \( \frac{2\phi}{3a}x+\frac{\phi}{2} \) \psi(x),
\end{aligned}
\end{equation}
where $\lambda=E^2-t_1^2-t_2^2-t_3^2$.
It is straightforward to see that the identification of the mass parameters is now given by
\begin{equation}
\begin{aligned}
m_1&=\( \frac{t_2 t_3}{t_1^2} \)^{2/3} e^{i\phi/3},\quad
m_2=\( \frac{t_3 t_1}{t_2^2} \)^{2/3} e^{i\phi/3},\quad
m_3=\( \frac{t_1 t_2}{t_3^2} \)^{2/3} e^{i\phi/3},\\
\cE'&=(t_1 t_2 t_3)^{-2/3} e^{i\phi/6} \lambda.
\end{aligned}
\end{equation}
For the contrast, we also consider the asymmetric triangular lattice whose Hamiltonian is given by
\begin{equation}
\begin{aligned}
H_\text{tri}=T_1(e^{i\x}+e^{-i\x})+T_2(e^{i\y}+e^{-i\y})+T_3(e^{i\x+i\y}+e^{-i\x-i\y}).
\end{aligned}
\end{equation}
For this model, the mass identification should be 
\begin{equation}
\begin{aligned}
m_1&=\( \frac{T_1^2}{T_2T_3} \)^{2/3},\qquad
m_2=\( \frac{T_2^2}{T_3T_1} \)^{2/3},\qquad
m_3=\( \frac{T_3^2}{T_1T_2} \)^{2/3}, \\
\cE'&=(T_1 T_2 T_3)^{-1/3} \cE.
\end{aligned}
\end{equation}

\section{An application: bandwidth in weak magnetic regime}\label{sec:application}
In the previous section, we find a relation between the Bloch electron on the honeycomb lattice and quantum geometry of local $\cB_3$.
In this section, we use this relation to compute the bandwidth of the electron spectrum in the weak magnetic regime.
Throughout this paper, we focus on the spectrum of $\lambda=E^2-3$ rather than the original energy $E$.
It is straightforward to translate the results here into $E$.

\subsection{Known results}
The spectrum in the weak magnetic regime was studied in great detail in \cite{hatsuda2018}.
Here we summarize the results in \cite{hatsuda2018}.
We are interested in the spectrum near the top $\lambda=6$ and the bottom $\lambda=-3$ in the weak limit $\phi \sim 0$.
The positions of the bands are approximately explained by the perturbative expansion of $\phi$:
\begin{equation}
\begin{aligned}
\lambda_\text{top}^\text{pert}(n,\phi)&=6-\sqrt{3}(2n+1)\phi+\frac{3n^2+3n+1}{6}\phi^2-\frac{n(2n^2+3n+1)}{36\sqrt{3}}\phi^3+\cO(\phi^4), \\
\lambda_\text{bot}^\text{pert}(n,\phi)&=-3+\sqrt{3} n \phi-\frac{n^2}{2} \phi^2
-\frac{n(n^2+2)}{18\sqrt{3}}\phi^3+\cO(\phi^4).
\end{aligned}
\label{eq:lambda-pert}
\end{equation}
where $n$ denotes the Landau level. The bottom spectrum for the lowest Landau level $n=0$ is very special. In this mode, there is no quantum correction.
%This fact implies an existence of a supersymmetric quantum mechanical structure \cite{ezawa2008}.
This fact suggests that there is a supersymmetry in this case; the existence of a supersymmetric quantum mechanical structure was already noticed in the continuum limit \cite{ezawa2008}.

The bandwidth is non-perturbative in the magnetic flux $\phi$. It is never visible in the perturbative expansions \eqref{eq:lambda-pert}.
One of the main results in \cite{hatsuda2018} is the detailed quantitative analysis of the non-perturbative bandwidth.
At the top of the spectrum, the leading non-perturbative contribution takes the form
\begin{equation}
\begin{aligned}
\Delta \lambda_\text{top}^\text{band}(n, 2\pi/Q)\approx \frac{108 \cdot 3^{1/4}}{n!} \biggl( \frac{6\sqrt{3}}{\pi} \biggr)^n Q^{n-\frac{1}{2}} e^{-\frac{S_0}{2\pi}Q}
\cP_\text{top}^\text{inst}(n,2\pi/Q), 
\end{aligned}
\label{eq:BW-top}
\end{equation}
where we have set $\phi=2\pi/Q$ ($Q \to \infty$), and non-perturbative magnitude $S_0$ is exactly given by
\begin{equation}
\begin{aligned}
S_0=\frac{2}{i}\int_{-\pi/2}^{\pi/2} dq \, \arccos\( \frac{2}{\cos q}-\cos q \)=10.149416064\cdots.
\label{eq:A}
\end{aligned}
\end{equation}
The function $\cP_\text{top}^\text{inst}(n,\phi)$ is the most non-trivial part, and its closed form is not known.
The careful numerical analysis in \cite{hatsuda2018} revealed its small $\phi$ expansion:
\begin{equation}
\begin{aligned}
\log \cP_\text{top}^\text{inst}(n,\phi)&=-\frac{6n^2+42n+19}{72\sqrt{3}}\phi-\frac{2n^3+15n^2+15n+6}{864}\phi^2 \\
&\quad-\frac{15n^4+138n^3+258n^2+297n+166}{46656\sqrt{3}}\phi^3+\cO(\phi^4) .
\end{aligned}
\end{equation}

The spectrum near the bottom edge is more involved. 
In this case, it was observed in \cite{hatsuda2018} that there is a pair of subbands for each Landau level $n \geq 1$ whose bandwidths are almost same.
The gap of these two subbands is almost regarded as a zero-gap. See table 1 in \cite{hatsuda2018}.
We distinguish these two subbands by subscript $\pm$.
At the leading order, their bandwidths have the same form:
\be
\Delta \lambda_\text{bot,$\pm$}^\text{band}(n, 2\pi/Q) \approx \frac{3^{\frac{3(n+1)}{2}}\sqrt{n}}{(2\pi)^{n-\frac{1}{2}}n!} Q^{n-1}e^{-\frac{S_0}{10\pi}Q}
\cP_\text{bot}^\text{inst}(n,2\pi/Q),
\qquad n \geq 1,
\label{eq:BW-bot}
\ee
where $S_0$ is the same number as \eqref{eq:A}. 
The function $\cP_\text{bot}^\text{inst}(n,\phi)$ is given by
\begin{equation}
\begin{aligned}
\log \cP_\text{bot}^\text{inst}(n,\phi)&=-\frac{30n^2+72n+11}{72\sqrt{3}}\phi-\frac{34n^3+96n^2+49n+16}{432}\phi^2 \\
&\quad -\frac{4470n^4+17280n^3+14910n^2+12960n+1081}{58320\sqrt{3}}\phi^3+\cO(\phi^4),
\end{aligned}
\end{equation}
As mentioned before, the lowest Landau level $n=0$ is special.
%As observed in \eqref{eq:lambda-pert}, the lowest perturbative spectrum $\lambda_\text{bot}^\text{pert}(0,\phi)=-3$ is protected from the quantum corrections.  It implies an existence of a supersymmetry.
Its band structure is quite different from the excited levels.
We do not look at it in this paper. See \cite{hatsuda2018} in detail.

\subsection{Relation to topological string free energy}
We should note that almost all the results in the previous subsection were guesses based on the thorough numerical study in \cite{hatsuda2018}.
There is no systematic way to compute or predict the higher order corrections to the unknown functions $\cP_\text{top}^\text{inst}(n,\phi)$ and $\cP_\text{bot}^\text{inst}(n,\phi)$.
In this subsection, we will relate these functions to the topological string free energy on the quantum local $\cB_3$ geometry.
Using this nice connection, we can predict the higher order corrections to $\cP_\text{top}^\text{inst}(n,\phi)$ and $\cP_\text{bot}^\text{inst}(n,\phi)$ by using the topological string technique.
The similar approach in the Hofstadter model is found in \cite{duan2019}.

Following \cite{duan2019}, we introduce new functions $\cA_\text{top}(n,\phi)$ and $\cA_\text{bot}(n,\phi)$ by
\begin{equation}
\begin{aligned}
\cP_\text{top}^\text{inst}(n,\phi)&=-\frac{1}{2\sqrt{3}\phi} \frac{\del \lambda_\text{top}^\text{pert}(n,\phi)}{\del n} e^{-\cA_\text{top}(n,\phi)}, \\
\cP_\text{bot}^\text{inst}(n,\phi)&=\frac{1}{\sqrt{3}\phi} \frac{\del \lambda_\text{bot}^\text{pert}(n,\phi)}{\del n} e^{-\cA_\text{bot}(n,\phi)}.
\end{aligned}
\label{eq:A-funcs}
\end{equation}
From the results in the previous section, one easily finds $\cA_\text{top}(n,\phi)$ and $\cA_\text{bot}(n,\phi)$ as a series expansion in $\phi$,
\begin{subequations}
\begin{align}
\cA_\text{top}\left( n=\nu-\frac{1}{2},\phi \right)
&=
\frac{-1+12\nu^2}{144 \sqrt{3}}\phi
+\frac{\nu(3+4\nu^2)}{1728} \phi ^2
+\frac{1051+1176\nu^2+240\nu^4}{746496 \sqrt{3}}\phi^3 \notag \\
&\quad+\frac{\nu(85813+41640 \nu ^2+2832 \nu ^4)}{149299200}\phi^4+ \cO(\phi^5),
\label{Atop}
\\
\cA_\text{bot}(n,\phi)
&=
\frac{11+30 n^2}{72 \sqrt{3}}\phi+\frac{n(49+34 n^2)}{432}\phi ^2+\frac{1081+14910 n^2+4470 n^4}{58320 \sqrt{3}}\phi^3\notag\\
&\quad+\frac{n(158387+441730n^2+72078n^4)}{2332800}\phi^4+\cO (\phi ^5 ).
\label{Abot}
\end{align}
\end{subequations}
It is claimed in \cite{duan2019} that these functions are related to the free energy of the refined topological string in the Nekrasov--Shatashvili limit at conifold singular points. 
This is natural because, as discussed in \cite{hatsuda2016}, the conifold singular points corresponds to the band edges.
Therefore it is expected that the expansion around the band edges are captured by the conifold frame.
Below, we refer to the free energy in the Nekrasov--Shatashvili limit as the NS free energy for short.
We will briefly review the refined topological string in appendix~\ref{RefTopStr}.

One of the main results in  \cite{duan2019} is the following relation between the NS free energy of local $\mathbb{F}_0$ in the conifold frame%
\footnote{The definition of $F_c (t_c,\hbar)$ in appendix~\ref{RefTopStr} is a little bit different from that in \cite{duan2019}, and this difference results in an additional term $-t_c$ in \eqref{relTopCond}.} $F_c (t_c,\hbar)$ and the function $\cA(n,\phi)$ in the Hofstadter model on square lattice,
\be\ba
\cA^\text{Hof}\left( n,\phi \right) =\biggl[\biggl[ \frac{1}{\hbar} \biggl(\frac{\del F_c^{\mathbb{F}_0}(t_c,\hbar)}{\del t_c}-t_c\biggr) \biggr]\biggr] \bigg|_{\hbar=-\phi,~ t_c = - \phi\nu},
\label{relTopCond}
\ea\ee
where $[[f(\hbar)]]$ denotes the power series of $f(\hbar)$ in $\hbar$ starting from $\cO(\hbar)$.
%where $A(\phi)$ is a constant determined by the classical B-period \eqref{PrepAndB} at conifold point that we will define later. We also denote $\left[ f(\phi) \right]$ the power series in $\phi$ starting from $\cO(\phi)$\footnote{By definition, we need not include the term $A/\phi$, however, this term plays an important role to find the relation in our case as we will see later.}.
Our goal is to find a similar relation between the local $\cB_3$ geometry and the honeycomb lattice. It turns out that we need to slightly modify the relation \eqref{relTopCond} in this case.

For our purpose, we need the NS free energy in the conifold frame. 
We will discuss how to compute it in appendix~\ref{RefTopStr}.
To fit the convention in the literature on the topological string theory, we slightly change the notation of the quantum mirror curve \eqref{eq:H-honey} as follows:
\begin{equation}
\begin{aligned}
H=e^{\xi}+e^{-\xi}+e^{\eta}+e^{-\eta}+e^{-\frac{i\hbar}{2}}e^{\xi+\eta}+e^{\frac{i\hbar}{2}} e^{-\xi-\eta}, \qquad [\xi,\eta]=i\hbar.
\end{aligned}
\end{equation}
where we have formally replaced $(i\x,i\y) \to (\xi,\eta)$. Under this replacement, the quantum parameters $\phi$ and $\hbar$ are related by $\hbar=-\phi$.
%In the current case, the band edges at $\phi=0$ are just $\lambda=6$ (top) and $\lambda=-3$ (bottom).
In general, the NS free energy has the following expansion:
\be\ba
F_c(t_c,\hbar) = \sum_{n\geq0} F_{c,n}(t_c) \hbar^{2n}.
\label{NSGenusExp}
\ea\ee
where the coefficients $F_{c,n}(t_c)$ have the mass dependence, and in our case they are related to the quantum parameter as in \eqref{eq:mass-honey}. Therefore one should keep in mind that $F_{c,n}(t_c)$ implicitly depend on $\hbar$.

%%%%%%%%%%%%%%%%%
\paragraph{Spectrum near the top}
\par~\\
Let us consider the spectrum near the top $\lambda=6$. 
Using the method in appendix~\ref{RefTopStr}, we obtain the NS free energy $F_{c}(t_c,\hbar)$ order by order.
See \eqref{eq:FNS-top}.
We further re-expand its coefficients $F_{c,n}(t_c)$ in terms of $\hbar$, and find the following results:
\begin{align}
F_{c,0}^{\text{top}}(t_c)
&=\frac{1}{2} t_{c}^2 \log \left(\frac{t_{c}}{12} \right)-\frac{t_{c}^2}{4}-\frac{t_{c}^3}{36}+\frac{t_{c}^4}{576}
+\left(-\frac{t_{c}^2}{216} +\frac{t_{c}^3}{1944} -\frac{t_{c}^4}{31104} \right) \hbar ^2 \notag
\\
&\quad
+\left(-\frac{5 t_{c}^2}{46656} +\frac{t_{c}^3}{34992} -\frac{131 t_{c}^4}{20155392} \right) \hbar ^4 + \cO(\hbar^6,t_{c}^5),
\notag\\ %%
F_{c,1}^{\text{top}}(t_c)
&=
-\frac{\log t_{c}}{72}-\frac{23 t_{c}}{432}+\frac{19 t_{c}^2}{3456}-\frac{97 t_{c}^3}{93312}+\frac{1201 t_{c}^4}{4976640}
\notag\\
&\quad
+\left(-\frac{25 t_{c}}{23328} +\frac{85 t_{c}^2}{186624} -\frac{199 t_{c}^3}{1119744}  +\frac{5287 t_{c}^4}{80621568} \right) \hbar ^2
\\
&\quad
+\left(-\frac{t_{c}}{46656} +\frac{2111 t_{c}^2}{120932352} -\frac{47 t_{c}^3}{4478976}  +\frac{30923 t_{c}^4}{5804752896} \right) \hbar ^4 + \cO(\hbar^6,t_{c}^5),
\notag\\
F_{c,2}^{\text{top}}(t_c)
&=
-\frac{7}{51840 t_{c}^2}+\frac{581 t_{c}}{2239488} -\frac{18187 t_{c}^2}{298598400}-\frac{47 t_{c}^3}{134369280}  + \frac{152191 t_{c}^4}{12899450880} 
\notag\\
&\quad
+\left(\frac{2639 t_{c}}{134369280}-\frac{84901 t_{c}^2}{4837294080}+\frac{24817 t_{c}^3}{2418647040} -\frac{361249 t_{c}^4}{77396705280}  \right) \hbar ^2
\notag\\
&\quad
+\left(\frac{991 t_{c}}{4837294080} -\frac{52645 t_{c}^2}{69657034752} +\frac{2066413 t_{c}^3}{2350924922880}  -\frac{1260497 t_{c}^4}{1857520926720}  \right) \hbar ^4
+ \cO(\hbar^6,t_{c}^5).\notag
\end{align}
%where we denote $t_{c}$ by the classical A-period \eqref{ClaAperiod} expressed as a function of $z_{c,1}$ through \eqref{ConiFrameTop}.
%Since the conifold point depends on $\hbar$, the classical A-period at the conifold point $A$ is now the function of $\hbar$, and has following expansion in $\hbar$,
It is also important to notice that we need to modify \eqref{eq:A} because the mirror curve itself depends on $\hbar$ thorough the mass parameters.
We find that it should be modified as
\be\ba
S^\text{top}(\hbar) 
&:= \frac{2}{i} \int_{-\pi/2}^{\pi/2} \rd q \arccos \left( \frac{3 \cos \left( \hbar/6 \right) }{2 \cos (q)} -  \frac{ \cos \left(2q - \hbar/2 \right) }{2 \cos (q)} \right) 
\\
&= S_0 - \frac{\hbar^2}{6\sqrt{3}} - \frac{\hbar^4}{972\sqrt{3}} + \cO(\hbar^6).
\ea\ee
By combining these results, and comparing with \eqref{Atop}, we find
\be\ba
\cA_\text{top}\left( n,\phi \right) =\biggl[\biggl[ -\frac{S^\text{top}(\hbar)}{\hbar} +  \frac{\sqrt{3}}{\hbar} \biggl(\frac{\del F_c^\text{top}(t_{c},\hbar)}{\del t_{c}}-t_c  \biggr) \biggr]\biggr] \bigg|_{\hbar=-\phi, t_{c} = - \frac{\phi\nu}{\sqrt{3}}}.
\label{AtopFNS}
\ea\ee
Therefore, we conclude that the instanton correction to the function $\cP_{\text{top}}^{\text{inst}}(n,\phi)$ can be computed from the topological string free energy.

%Here we have focused on the non-trivial function $\cA_\text{top}\left( n,\phi \right)$. 
It is interesting to note that the right hand side in \eqref{AtopFNS} also reproduces the prefactor in \eqref{eq:BW-top}.
Let us see it in detail.
One finds
\begin{align}
-\frac{S^\text{top}(\hbar)}{\hbar} +  \frac{\sqrt{3}}{\hbar} \biggl(\frac{\del F_c^\text{top}(t_{c},\hbar)}{\del t_{c}}-t_c\biggr)
&=\frac{S_0}{\phi}+\cA_\text{top}(n,\phi)+\nu \log \( -\frac{\nu \phi}{12\sqrt{3}}\)\\
&\quad-\nu-\frac{1}{12\nu}+\frac{7}{2880\nu^3}-\frac{31}{40320\nu^5}+\frac{127}{215040\nu^7}+\cdots. \notag
\end{align}
where we have abbreviated the identification $\hbar=-\phi$, $t_{c} = - \phi\nu/\sqrt{3}$.
It is easy to guess that the infinite sum on the right hand side is related to the asymptotic expansion of the gamma function:
\begin{equation}
\begin{aligned}
-\nu-\frac{1}{12\nu}+\frac{7}{2880\nu^3}-\frac{31}{40320\nu^5}+\frac{127}{215040\nu^7}+\cdots
=\log \Gamma\(\nu+\frac{1}{2} \)-\nu \log \nu-\frac{\log(2\pi)}{2}.
\end{aligned}
\end{equation}
Therefore we obtain
\begin{equation}
\begin{aligned}
\exp\biggl[ \frac{S^\text{top}(\hbar)}{\hbar}- \frac{\sqrt{3}}{\hbar} \biggl(\frac{\del F_c^\text{top}(t_{c},\hbar)}{\del t_{c}}-t_c\biggr)\biggr]
=\frac{\sqrt{2\pi}}{\Gamma(\nu+\frac{1}{2})}\biggl( -\frac{12\sqrt{3}}{\phi} \biggr)^\nu e^{-\frac{S_0}{\phi}-\cA_\text{top}(n,\phi)}.
\end{aligned}
\end{equation}
We compare this result with the leading bandwidth \eqref{eq:BW-top}. Using \eqref{eq:A-funcs}, it is rewritten as
\begin{equation}
\begin{aligned}
\Delta \lambda_\text{top}^\text{band}\approx -\frac{9}{\sqrt{2\pi} n!} \biggl( \frac{12\sqrt{3}}{\phi} \biggr)^{n+\frac{1}{2}} \frac{\del \lambda_\text{top}^\text{pert}(n,\phi)}{\del n} e^{-\frac{S_0}{\phi}-\cA_\text{top}(n,\phi)}.
\end{aligned}
\end{equation}
Recalling $\nu=n+1/2$, these two results match up to a numerical factor. We conclude that the leading bandwidth is expressed as
\begin{equation}
\begin{aligned}
 \Delta \lambda_\text{top}^\text{band} \approx C_\text{top} \frac{ \del \lambda_\text{top}^\text{pert}}{\del n} \exp\biggl[ \frac{S^\text{top}(\hbar)}{\hbar}- \frac{\sqrt{3}}{\hbar} \biggl(\frac{\del F_c^\text{top}(t_{c},\hbar)}{\del t_{c}}-t_c\biggr)\biggr],
\end{aligned}
\end{equation}
where $C_\text{top}$ is a numerical constant.

%%%%%%%%%%%%%%%%%
\paragraph{Spectrum near the bottom}
\par~\\
The computation of the NS free energy corresponding to the spectrum near the bottom is almost the same. 
In this case, we have the following expansion coefficients:
\begin{align}
F^{\text{bot}}_{c,0}(t_{c})
&=
\left( \frac{t_{c}^2 \log \hbar}{4}  -\frac{t_{c}^3}{4 \sqrt{3} \hbar} -\frac{t_{c}^4}{16 \hbar^2}  \right)
+\left(\frac{t_c^2}{4} \log \left(\frac{\sqrt{3} t_c}{27}  \right)  -\frac{3t_{c}^2}{8}  -\frac{5 t_{c}^3}{36}  +\frac{17 t_{c}^4}{288}  \right)
\notag\\
&\quad
+\left(\frac{5 t_{c}^2}{24 \sqrt{3}} -\frac{17 t_{c}^3}{144 \sqrt{3}} +\frac{149 t_{c}^4}{1296 \sqrt{3}} \right) \hbar
+\left(\frac{19 t_{c}^2}{864}-\frac{61 t_{c}^3}{1944}+\frac{15559 t_{c}^4}{311040}\right) \hbar^2
+ \cO(t_c^5,\hbar^3),
\notag\\
F^{\text{bot}}_{c,1}(t_{c})
&=
\left(
- \frac{\log \left( -27 \sqrt{3} \hbar \right)}{144} +\frac{t_{c}}{48 \sqrt{3}  \hbar}   + \frac{t_{c}^2}{96 \hbar^2}    +\frac{t_{c}^3}{48 \sqrt{3} \hbar^3}   
 \right)
 \notag \\
&\quad
+\left(
- \frac{ \log  t_{c}}{144}  -\frac{31 t_{c}}{432}  +\frac{37 t_{c}^2}{576}    -\frac{1075 t_{c}^3}{11664}  
\right)
%\\
%&\qquad
% - \left( \frac{3}{64 \hbar^4} + \frac{10363}{23040}\right)  t_{c}^4
% \biggr\}
\notag\\
&\quad
+\left(\frac{31}{864\sqrt{3}}-\frac{37 t_{c}}{576 \sqrt{3}}+\frac{1075 t_{c}^2}{7776 \sqrt{3}}-\frac{10363 t_{c}^3}{34560 \sqrt{3}} %-\frac{60971 t_{c}^4}{31104 \sqrt{3}} 
\right) \hbar
\label{FcBot}\\
&\quad
+\left(\frac{125}{31104}-\frac{419 t_{c}}{23328}+\frac{117881 t_{c}^2}{1866240}-\frac{26803 t_{c}^3}{139968} %-\frac{75774245 t_{c}^4}{47029248} 
\right) \hbar^2 + \cO(t_{c}^4,\hbar^3),
\notag\\
F^{\text{bot}}_{c,2}(t_{c})
&=
\left( 
  -\frac{7  t_c}{5760 \sqrt{3} \hbar ^3}  -\frac{7  t_c^2}{3840 \hbar ^4}   -\frac{7 t_c^3 }{960  \sqrt{3} \hbar ^5}  
\right)
\notag\\
&\quad
+\left( 
-\frac{7}{103680 t_c^2}  -\frac{15443  t_c}{1399680}    +\frac{1693691 t_c^2}{37324800}   -\frac{260975 t_c^3}{1679616}    
\right)
\notag\\
&\quad
 %+ \left(-\frac{21}{256 \hbar ^6}+\frac{676441083}{156255350}\right)  t_c^4
%\\
%&\quad
+\left(
-\frac{1693691 t_c}{37324800 \sqrt{3}}+\frac{260975 t_c^2}{1119744 \sqrt{3}}-\frac{1086114361 t_c^3}{3386992574\sqrt{3}} %+\frac{196103399220 t_c^4}{6241645229\sqrt{3}}
\right)\hbar
\notag\\
&\quad
+\left(
-\frac{182129 t_c}{5598720}+\frac{255509179 t_c^2}{1209323520}-\frac{3922260391 t_c^3}{3733110103} %+\frac{997540275 t_c^4}{24879343}
\right)\hbar^2
+\cO(t_c^4, \hbar^3).\notag
\end{align}
%As is the same in the previous case, we denote $t_{c}$ by the classical A-period \eqref{ClaAperiod} expressed as a function of $z_{c,2}$ through \eqref{ConiFrameBot}.
This expression includes the negative power corrections in $\hbar$. However, after setting $t_{c} = - \phi \nu/\sqrt{3}$ and $\hbar = -\phi$, all of them disappear. By comparing \eqref{FcBot} with \eqref{Abot}, we find following relation,
\be\ba
\cA_\text{bot}\left( n,\phi \right) =\biggl[\biggl[ -\frac{S^\text{bot}(\hbar)}{\hbar} +  \frac{\sqrt{3}}{\hbar} \biggl(\frac{\del F_c^\text{bot}(t_{c},\hbar)}{\del t_{c}}-t_c  \biggr)\biggr]\biggr] \bigg|_{\hbar=-\phi, t_{c} = - \frac{\phi\nu}{\sqrt{3}}}.
\label{AvsFatBot}
\ea\ee
where
\be\ba
S^\text{bot}(\hbar)
&=- \frac{1}{i} \int_{-\pi/2}^{\pi/2} \rd q \arccos \left( \frac{3-6\cos(\hbar/3)}{8\sin (\pi/6-\hbar/6) \cos q} -  \frac{ \cos \left(2q - \hbar/2 \right) }{2 \cos (q)} \right) 
\\
& =\frac{S_0}{5}-\pi^2 i -\frac{\hbar \log \hbar}{2}+\( \frac{1}{2}+\frac{3\log 3}{4}\) \hbar + \frac{\hbar^2}{24\sqrt{3}} - \frac{\hbar^3}{432} +\frac{\hbar^4}{3888\sqrt{3}}+ \cO(\hbar^5).
%- \frac{ \hbar^4}{3888\sqrt{3}} + \cO(\hbar^5).
\label{AbotFNS}
\ea\ee
In this case, the relation to the bandwidth \eqref{eq:BW-bot} is much more involved than the top case.
We briefly sketch the computation based on some guesses.
We have
\begin{equation}
\begin{aligned}
-\frac{S^\text{bot}(\hbar)}{\hbar} +  \frac{\sqrt{3}}{\hbar} \biggl(\frac{\del F_c^\text{bot}(t_{c},\hbar)}{\del t_{c}}-t_c  \biggr)
=\frac{S_0}{5\phi}+\cA_\text{bot}(n,\phi)-\frac{\pi^2 i}{\phi}+\frac{\log(-\phi)}{2}\\
-\frac{1}{2}-\frac{3\log 3}{4}+\frac{\nu}{2} \log\( \frac{\nu \phi^2}{27}\)-\nu+\sum_{n \geq 0} s_n(\nu),
\end{aligned}
\end{equation}
where $s_n(\nu)$ is an infinite sum of $\nu$ coming from $F_{c,n}^\text{bot}(t_c)$.
Using \eqref{FcBot}, we observe
\begin{equation}
\begin{aligned}
s_0(\nu)&=-\frac{\nu}{2}-\frac{\nu^2}{4}-\frac{\nu^3}{12}-\frac{\nu^4}{24}+\cdots=-\nu-\frac{1}{2}(1-\nu)\log(1-\nu), \\
s_1(\nu)&=-\frac{1}{48\nu}+\frac{1}{48}+\frac{\nu}{48}+\frac{\nu^2}{48}+\cdots=-\frac{1}{48\nu}-\frac{1}{48(\nu-1)}, \\
s_2(\nu)&=\frac{7}{5760\nu^3}-\frac{7}{5760}-\frac{7\nu}{1920}-\frac{7\nu^2}{960}+\cdots
=\frac{7}{5760\nu^3}+\frac{7}{5760(\nu-1)^3}
\end{aligned}
\end{equation}
We further guess that these are resummed to the gamma function:
\begin{align}
-\frac{1}{48\nu}+\frac{7}{5760\nu^3}+\cdots&=\frac{1}{2}\log \Gamma\( \nu+\frac{1}{2} \)-\frac{\nu}{2}(\log \nu-1)-\frac{\log(2\pi)}{4}, \\
-\frac{1}{48(\nu-1)}+\frac{7}{5760(\nu-1)^3}+\cdots&=\frac{1}{2}\log \Gamma\( \nu-\frac{1}{2} \)-\frac{\nu-1}{2}(\log( \nu-1)-1)-\frac{\log(2\pi)}{4} \notag
\end{align}
Under these assumptions, we observe that the leading bandwidth \eqref{eq:BW-bot} is reproduced by
\begin{equation}
\begin{aligned}
 \Delta \lambda_{\text{bot},\pm}^\text{band} \approx C_\text{bot} \frac{ \del \lambda_{\text{bot},\pm}^\text{pert}}{\del n} \exp\biggl[ \frac{S^\text{bot}(\hbar)}{\hbar}- \frac{\sqrt{3}}{\hbar} \biggl(\frac{\del F_c^\text{bot}(t_{c},\hbar)}{\del t_{c}}-t_c\biggr)\biggr],
\end{aligned}
\end{equation}
where $C_\text{bot}$ is a numerical factor.

\section{Conclusion}\label{sec:con}
In this paper we proposed a new connection between the honeycomb lattice model and topological string theory. It is a non-trivial generalization of the original proposal in \cite{hatsuda2016, duan2019}.
%We found the exact relation between topological string on local $\cB_3$ with unconventional moduli identification and the Hofstadter model on the honeycomb lattice.
The non-perturbative corrections to the spectrum near the top or the bottom can be expressed by the NS free energy on local $\cB_3$ geometry.
This connection allows us to predict the higher order corrections to the function $\cP^\text{inst}(n,\phi)$ systematically.

We would like to note that the local $\cB_3$ with $m_1=m_2=m_3=1$ in \eqref{eq:mirror-B3} describes the Hofstadter model on the triangular lattice discussed in \cite{hatsuda2017}. Then, we expect the relations \eqref{AtopFNS} and \eqref{AbotFNS} to be satisfied for this case too. Actually, we have checked that the similar relations hold by replacing the NS free energies $F_c^{\text{top/bot}}(t_{c})$ and instanton actions $A_{\text{top/bot}}(\hbar)$ with those for $m_1=m_2=m_3=1$\footnote{We thank Zhaojie Xu for calculating $A_{\text{bot}}(\hbar)$.}.

As a further generalization, it would be interesting to consider non-hermitian cases. The non-hermitian Hofstadter model was discussed in \cite{matveenko2014, chernodub2015}. Na\"ively, the model would correspond to the topological string on genus-zero mirror curve. However, the (quantum) A-period is trivial for this curve. Therefore, firstly one need to reconsider what mirror curve corresponds to. %Or, we need to find how to connect the Hofstadter model to the genus-zero mirror curve.

Since there are various kinds of the mirror curves in the topological string side, we can investigate the branch cut of the quantum A-period which corresponds to the band spectrum in the Hofstadter model, if it exists. In this sense, we can give a lot of predictions from the topological string side. Especially, it would be interesting to find the Hofstadter model corresponding to the higher genus mirror curve. Even in this case the topological string would be powerful method to study the Hofstadter model systematically.

\section*{Acknowledgement}
We would like to thank Minxin Huang for valuable discussions.
The work of YH is supported by JSPS KAKENHI Grant Number JP18K03657.
The work of YS is supported by a grant from the NSF of China with Grant No: 11947301.

\appendix

\section{Refined topological string and NS limit}\label{RefTopStr}

Here we briefly review the refined topological string. Originally, the refined topological string is proposed in \cite{iqbal2009} to generalize the geometric engineering \cite{katz1997, katz1997a, dijkgraaf2002, hollowood2008}. The free energy of the refined topological string is given by
\be\ba
F(t,\epsilon_1,\epsilon_2) = \sum_{g,n\geq0}(\epsilon_1 + \epsilon_2)^{2n} (\epsilon_1 \epsilon_2)^{g-1} F_{n,g}(t),
\ea\ee
where $\epsilon_{1,2}$ are two deformation parameters.
Through the geometric engineering, the partition function defined by $Z=\re^{F(t,\epsilon_1,\epsilon_2)}$ agrees with the Nekrasov partition function \cite{nekrasov2003} of 5d $\cN=1$ gauge theory with several gauge groups.
In the unrefined limit $\epsilon_1+\epsilon_2=0$, the refined topological string reduces to the usual topological string,
which is determined by $F_{0,g}(t)$.

Another interesting limit is the Nekrasov--Shatashvili limit defined by turning off one of the omega deformation parameters \cite{nekrasov2010},
\be\ba
F(t, \hbar) = \lim_{\epsilon_2 \to 0} \epsilon_1 \epsilon_2 F(t,\epsilon_1,\epsilon_2)|_{\epsilon_1=\hbar}=\sum_{n=0}^\infty \hbar^{2n}F_{n,0}(t).
\ea\ee
We call this free energy the NS free energy for short.
Also, we denote $F_{n,0}(t)$ by $F_{n}(t)$.
One of the method to calculate the NS free energy is to solve the refined holomorphic anomaly equation \cite{huang2012} that is a generalization of the holomorphic anomaly equation \cite{bershadsky1994}. 
The refined holomorphic anomaly equation is the recursive equation for $F_{n,g}(t)$.
Using it, we can obtain the NS free energy. 
The explicit computation has been done in e.g. \cite{codesido2018a, codesido2019}.
In this paper, we instead utilize the operator method discussed in  \cite{mironov2010, huang2012b} as a more efficient way to calculate the NS free energy.

\subsection{Solving the Picard--Fuchs equation}\label{PFsol}
To obtain the NS free energy, we first compute the classical periods by solving the Picard--Fuchs (PF) equation.
The PF equation for the mirror curve \eqref{eq:mirror-B3} with $m_1=m_2=m_3=\re^{-i \hbar/6}$ is given by
\be\ba
\left[d_3(z) \theta_z^3 + d_2(z) \theta_z^2 + d_1(z) \theta_z \right] w(z) =0,
\label{PFlarge}
\ea\ee
where $z=1/\lambda$ and $\theta_z=z \frac{d}{dz}$. The coefficients are given by
\be\ba
&d_1(z) = - 6 z^2 (2+3c_3 z)^3,
\\
&d_2(z)= -z(c_3+108z+405 c_3 z^2 +540 c_3^2 z^3 +243 c_3^3 z^4),
\\
&d_3(z)= (1+c_3 z)(2+3c_3 z)(1-27 z^2-27 c_3 z^3),
\ea\ee
where
\begin{equation}
\begin{aligned}
c_n:=e^{\frac{in\hbar}{6}}+e^{-\frac{in\hbar}{6}}=2\cos \frac{n\hbar}{6}.
\end{aligned}
\end{equation}
Since redefining $u(z)=\theta_z w(z)$, the PF equation reduces to the second order ODE, we have essentially two independent solutions.
These are called the classical A- and B-periods.
The PF equation \eqref{PFlarge} has regular singular points at
\begin{equation}
\begin{aligned}
z=0,\;\;-\frac{1}{2\cos \frac{\hbar}{2}},\;\;-\frac{1}{3\cos\frac{\hbar}{2}},\;\;\frac{1}{6\cos \frac{\hbar}{6}},\;\;\frac{2\sin(\frac{\pi}{6}\pm \frac{\hbar}{6})}{3-6\cos\frac{\hbar}{3}},\;\;\infty.
\end{aligned}
\end{equation}

\paragraph{Solutions corresponding to the top edge}
\par~\\
Let us consider the top edge $\lambda=6$.
The only candidate is the singularity at
\begin{equation}
\begin{aligned}
z=\frac{1}{3c_1},\qquad c_1=2\cos \frac{\hbar}{6}.
\end{aligned}
\end{equation}
It is convenient to change the variable as follows:
\begin{equation}
\begin{aligned}
z_c=1-3c_1 z.
\end{aligned}
\end{equation}
We can easily construct the local solutions to the PF equation around $z_c=0$ by the Frobenius method.

One of the solutions has the non-logarithmic behavior that is called the A-period:
\begin{equation}
\begin{aligned}
w_{c,0}^A(z_c)=z_c+\frac{5c_1^2-2}{6c_1^2}z_c^2+\frac{19c_1^4-14c_1^2+10}{27c_1^4}z_c^3+\cO(z_c^4)
\end{aligned}
\end{equation}
The other called the classical B-period has the logarithmic term:
\be\ba
w_{c,0}^B(z_c)=w_{c,0}^A(z_c)\log\( \frac{c_1^2-1}{9c_1^2}z_c \)+\widetilde{w}_{c,0}^B(z_c)
\ea\ee
where
\be\ba
\widetilde{w}_{c,0}^B(z_c)=\frac{3 c_1^4-5 c_1^2+4}{4 c_1^2 (c_1^2-1)}z_c^2+\frac{148 c_1^8-460 c_1^6+732 c_1^4-658 c_1^2+211}{162 c_1^4 (c_1^2-1)^2}z_c^3+\cO(z_c^4)
\ea\ee
The prepotential near this singularity is then defined by
\begin{equation}
\begin{aligned}
t_c=w_{c,0}^A(z_c), \qquad
\frac{\del F_{c,0}^\text{top}(t_c)}{\del t_c}=w_{c,0}^B(z_c)
\end{aligned}
\label{eq:prepot}
\end{equation}
Eliminating $z_c$ from these two equations, we obtain the prepotential $F_{c,0}^\text{top}(t_c)$ as a function $t_c$.

\paragraph{Solutions corresponding to the bottom edge}
\par~\\
The argument of the bottom edge $\lambda=-3$ is more subtle.
In this case, there are three candidates of the singular points:
\begin{equation}
\begin{aligned}
z=-\frac{1}{3\cos\frac{\hbar}{2}},\;\;\frac{2\sin(\frac{\pi}{6}\pm \frac{\hbar}{6})}{3-6\cos\frac{\hbar}{3}}.
\end{aligned}
\end{equation}
All of these reduce to $z=-1/3$ in $\hbar \to 0$.
There seem to be no clear criteria how to choose one of them.
As a result of trial and error,%
\footnote{One non-trivial test is to impose a quantization condition $t_c=\hbar(n+1/2)/\sqrt{3}$ with \eqref{eq:quantum-geometry}. This quantization condition leads to the perturbative expansion of $\lambda$. We have confirmed that the quantum period near the singularity \eqref{eq:bottom-sing} correctly reproduces the bottom spectrum in \eqref{eq:lambda-pert}.}
 we found that the correct singularity that describes physics near the bottom edge is
\begin{equation}
\begin{aligned}
z=\frac{2\sin(\frac{\pi}{6}- \frac{\hbar}{6})}{3-6\cos\frac{\hbar}{3}}=\frac{c_1-\sqrt{3}s_1}{3(1-c_2)}.
\end{aligned}
\label{eq:bottom-sing}
\end{equation}
where $s_n:=2\sin(n\hbar/6)$.
So far, we have no idea whether the other two singular points play some physical roles in the 2d electron system.
We leave it for a future problem.
We can construct the local solutions around this singularity as well.
Let us define
\begin{equation}
\begin{aligned}
z_c=1-\frac{3(1-c_2)}{c_1-\sqrt{3}s_1}z
\end{aligned}
\end{equation}
The result is as follows:
\be\ba
w_{c,0}^B(z_{c}) =\frac{1}{2}  w_{c,0}^A(z_{c}) \log\left(\frac{-2 c_{4}+3c_2 -2 -\sqrt{3} s_2}{18(1-c_2)^2}z_c \right) + \widetilde{w}_{c,0}^B(z_{c})
\ea\ee
where
\be
\scalebox{0.85}{$\displaystyle
\ba
&w_{c,0}^A (z_{c})  
= z_{c} + \frac{5c_4-9 c_2+11- \sqrt{3} s_2}{6(1-c_2)^2} z^2_{c} 
\\
&\qquad\qquad\qquad
+ \frac{-240c_2 + 143 c_4 -69 c_6 +19 c_8 +\sqrt{3}\left(6s_2  +9s_4  -7s_6 \right)+309}{27(1-c_2)^4} z^3_{c,2}  
%\\
%&\qquad\qquad\qquad
+ \cO(z_{c,2}^4),
\\
& \tilde{w}_{c,2}^B (z_{c,2},\hbar) 
= \frac{-10c_2 +3 c_4 +c_6+3c_8+\sqrt{3}\left(-2s_2 -3s_4 -s_6 \right) - \frac{3\sqrt{3}c_1}{s_1} -6}{8(1+c_4)^2 } z_{c,2}^2
\\
&\qquad\qquad\qquad
-\frac{(1+c_2)^2}{648 s_1^2 (1+c_4)^4}
\\
&\qquad\qquad\qquad\quad
\times [
9708 c_2 -7754 c_4 +8085 c_6 -4348 c_8+1720 c_{10}-1020 c_{12} + 296 c_{14}
\\
&\qquad\qquad\qquad\quad\qquad
+\sqrt{3}\left(-978s_2 - 786s_4 +211 s_6 +696 s_8  +72 s_{10}  -164 s_{12} \right)
-13212
] z_{c,2}^3
\\
&\qquad\qquad\qquad
+ \cO(z_{c,2}^4).
\ea
$}
\ee
The prepotential $F_{c,0}^\text{bot}(t_c)$ is also computed by \eqref{eq:prepot}.

\subsection{Quantum periods}
Now we proceed to the NS free energy. 
The NS free energy is computed by the quantum periods.
Here we utilize the operator method proposed in \cite{mironov2010} and developed in \cite{huang2012b}. 
The classical periods receive the quantum corrections as
\begin{equation}
\begin{aligned}
w_c^I(z,\hbar)=\sum_{n=0}^\infty \hbar^{2n} w_{c,n}^I(z),\qquad I=A, B.
\end{aligned}
\end{equation}
The quantum corrections $w_{c,n}^I(z)$ can be obtained by acting differential operators \cite{mironov2010}:%
\footnote{The existence of such differential operators seems somewhat mysterious to the authors. To our knowledge, there seem no rigorous proofs of it. For Schr\"odinger's differential equations, the WKB solutions admit the existence of the operators \cite{mironov2010}.
In our case, we have difference equations rather than differential equations, but the WKB solutions still work. Therefore we expect that such differential operators widely exist even in difference equation.} 
\begin{equation}
\begin{aligned}
w_{c,n}^I(z)=\cD_n w_{c,0}^I(z).
\end{aligned}
\end{equation}
For local $\cB_3$, we do not find explicit forms of such operators, and we have computed it by ourselves.
We found the following explicit forms of the differential operators:
\begin{equation}
\begin{aligned}
\cD_1&=\frac{z^2}{2}\theta_z-\frac{2+c_3z-36z^2-36c_3z^3}{24(2+3c_3z)}\theta_z^2, \\
\cD_2&=-\frac{19z^2}{720}\theta_z-\frac{z^2(157+162c_3z+45c_3^2z^2)}{1440(2+3c_3 z)}\theta_z^2\\
&\quad
+\frac{23-23c_3z-369z^2-306c_3z^3-405c_3^2z^4}{8640(2+3c_3z)}\theta_z^3\\
&\quad-\frac{4-17c_3z-234z^2+36c_3z^3+270c_3^2z^4}{17280(2+3c_3z)}\theta_z^4.
\end{aligned}
\end{equation}
Note that here the operator $\cD_2$ contains $\theta_z^3$ and $\theta_z^4$, but by using the Picard--Fuchs equation \eqref{PFlarge} annihilating the classical period, all of the differential operators can be given by the linear combination of the first and second derivatives, 
\be\ba
\cD_n = a_{n}(z) \theta_z + b_{n}(z) \theta_z^2,
\ea\ee
where $a_{n}(z)$ and $b_{n}(z)$ are the rational functions of $z$ with the parameter $c_3$.
Since $a_{n}(z)$ and $b_{n}(z)$ are tedious long functions for $n \geq 2$, we do not write down their explicit forms in this paper.

The NS free energy in the conifold frame is finally given by
\begin{equation}
\begin{aligned}
t_c&=w_c^A(z_c,\hbar),\qquad
\frac{\del F_c(t_c,\hbar)}{\del t_c}&=w_c^B(z_c, \hbar).
\end{aligned}
\label{eq:quantum-geometry}
\end{equation}
By eliminating $z_c$ from these two equations, we obtain $F_{c,n}(t_c)$ as a function $t_c$.

\paragraph{The top edge}
\par~\\
We show a few corrections explicitly. The NS free energy corresponding to the top is as follows:
\begin{equation}
\scalebox{0.8}{$\displaystyle
\begin{aligned}
F_{c,0}^{\text{top}}(t_c)&=\frac{t_c^2}{2}\log \( \frac{c_1^2-1}{9c_1^2}t_c \)-\frac{t_c^2}{4}-\frac{c_1^4+c_1^2-8}{36 c_1^2 (c_1^2-1)}t_c^3
%\\
%&\quad
+\frac{c_1^8+2 c_1^6+129 c_1^4-520 c_1^2+280}{2592 c_1^4 (c_1^2-1)^2}t_c^4+\cO(t_c^5), %-\frac{529 t_c^6}{566870400} 
\\
F_{c,1}^{\text{top}}(t_c)&=-\frac{1}{72}\log t_c+\frac{-7 c_1^4-55 c_1^2+56}{432 c_1^2 (c_1^2-1)}t_c
-\frac{c_1^8+290 c_1^6-3039 c_1^4+4952 c_1^2-2312}{15552 c_1^4(c_1^2-1)^2}t_c^2\\
&\quad+\frac{19 c_1^{12}-186 c_1^{10}+2274 c_1^8-14393 c_1^6+30408 c_1^4-26484 c_1^2+8200}{34992 c_1^6 (c_1^2-1)^3}t_c^3+\cO(t_c^4), \\ %-\frac{16763 t_c^6}{257132413440}
F_{c,2}^{\text{top}}(t_c)&=-\frac{7}{51840 t_c^2}+\frac{1}{12597120 c_1^6 (c_1^2-1)^3}(1169 c_1^{12}-61014 c_1^{10}+596982 c_1^8\\
&\quad-2072275 c_1^6+3242280 c_1^4-2345700 c_1^2+641960)t_c+\cO(t_c^2).
\end{aligned}
$}
\label{eq:FNS-top}
\end{equation}

\paragraph{The bottom edge}
\par~\\
The NS free energy corresponding to the bottom is as follows:
\be
\scalebox{0.8}{$\displaystyle
\ba
F_{c,0}^{\text{bot}}(t_c)
&=
\frac{1}{8} \biggl[-1+2 \log\left(\frac{2-3c_2 +2c_4 -\sqrt{3}s_2}{18(1-c_2)^2}t_c \right) \biggr] t_c ^2 
\\
&\qquad
+\frac{-22 c_2  -7c_4  +c_6  -c_8 +\sqrt{3}\left(-2s_2  -5s_4   -s_6 \right) -\frac{9\sqrt{3}c_1}{s_1} -32}{72(1+c_4)^2}t_c^3
+ \cO(t_c^4),
%\\
%&\qquad
%+\frac{1}{5184(1-c_2)^4 (1+c_2)^2} 
%\\
%&\qquad\qquad\times
%\{-1720 c_2  -22 c_4 +672c_6  +55c_8  +4c_{10} - c_{12}
%\\
%&\qquad\qquad\qquad
%+\sqrt{3}\left(-698s_2  -916s_4 -138s_6  +68s_8  -2 s_{10} \right) +\frac{324}{s_1^2}  -2269
%\}t_c^4+ \cO(t_c^5),
 \\ %%
F_{c,1}^{\text{bot}}(t_c)
&=-\frac{1}{48}\log t_c 
-
\frac{104c_2  -38c_4  +7c_6  +\sqrt{3}\left(28 s_2  +31s_4 \right)  -\frac{3\sqrt{3}c_1}{s_1}+40}{288(1-c_2)^2 (1+c_2)}t_c
 \\
&\quad
-\frac{1}{10368(1-c_2)^4(1+c_2)^2}
 \\
&\quad\quad\times
\biggl[
-6776 c_2  -122c_4 +6672c_6 +1817 c_8 -148c_{10}  + c_{12}
 \\
&\qquad\qquad
+\sqrt{3}(-6934s_2  -5564 s_4  -2310 s_6 +1228 s_8 +146 s_{10} ) 
-\frac{324}{s_1^2}  -19331
\biggr] t_c^2
+\cO(t_c^3),
 \\
F_{c,2}^{\text{bot}}(t_c)
&=-\frac{7}{103680t_c^2} 
-\frac{1}{100776960(1-c_2)^6 (1+c_2)^3 s_1^3}
 \\
&\quad\quad\times
\biggl[
\sqrt{3}(36049794 c_1 -9245646 c_3 - 33028425 c_5  -13408014 c_7  +12987351 c_9  
 \\
&\qquad\qquad\quad
+ 10135515 c_{11} + 154503 c_{13}  -4170357 c_{15} +446112 c_{17}  + 99579 c_{19} )
 \\
&\qquad\qquad
-48254055 s_1  -62011576 s_3 -13583802 s_5 + 44316234 s_7 + 20562037 s_9 
 \\
&\qquad\qquad
- 4960998 s_{11}   - 13675827 s_{13} + 561876 s_{15} + 2027523 s_{17} - 209997 s_{19} +  18403 s_{21}
\biggr] t_c
 \\
&\qquad
+ \cO(t_c^2).
\ea
$}
\ee

\bibliographystyle{amsmod}
\bibliography{Hofstadter}

\end{document}